\documentclass[aps,pre,one,11pt,showpacs]{revtex4}%
\usepackage{epsfig}
\usepackage{graphicx}%
\usepackage{amsmath}%
\usepackage{amsfonts}%
\usepackage{amssymb}
\begin{document}
\title{The rheology of hard sphere suspensions at arbitrary volume fractions: An
improved differential viscosity model}
\author{Carlos I. Mendoza$^{1}$ and I. Santamar\'{\i}a-Holek$^{2}$ }
\affiliation{$^{1}$Instituto de Investigaciones en Materiales, Universidad Nacional
Aut\'{o}noma de M\'{e}xico, Apdo. Postal 70-360, 04510 M\'{e}xico, D.F., Mexico}
\affiliation{$^{2}$Facultad de Ciencias, Universidad Nacional Aut\'{o}noma de M\'{e}xico,
Circuito Exterior de Ciudad Universitaria, 04510, M\'{e}xico D. F., Mexico}

\begin{abstract}
We propose a simple and general model accounting for the dependence of the
viscosity of a hard sphere suspension at arbitrary volume fractions. The model
constitutes a continuum-medium description based on a recursive-differential
method that assumes a hierarchy of relaxation times. Geometrical information
of the system is introduced through an effective volume fraction that
approaches the usual filling fraction at low concentrations and becomes one at
maximum packing. The agreement of our expression for the viscosity with
experiments at low- and high-shear rates and in the high-frequency limit is
remarkable for all volume fractions.

\end{abstract}
\maketitle

%\pacs{}

\section{Introduction}

A large amount of work has been devoted to the understanding of the rheology
of suspensions due to its ubiquitousness and the central role that they play
in many technological processes. To control the structure and flow properties
of such suspensions is of crucial practical importance. Due to its many body
character, the study of the rheology of these systems is extremely difficult,
even the simplest suspension, composed of identical hard spheres, show rather
complex rheological phenomena. Since Einstein's calculation of the viscosity
for a dilute suspension of spheres \cite{einstein}, a lot of effort has been
devoted to the extension of Einstein's result to more concentrated suspensions
in which correlations among particles must be taken into account. However, in
spite of the remarkable works which make central contributions to the
understanding of the rheology of concentrated suspensions \cite{saito}%
-\cite{verberg}, the problem still constitutes a challenge for theoretical
description \cite{lionberger}. Essentially, most of these works introduce
particle correlations by taking into account hydrodynamic interactions, and
provide a conceptual framework that explains how these interactions modify the
behavior of the viscosity as a function of the volume fraction $\phi$ for
arbitrary values of the wave vector and the frequency: $\eta(k,\omega;\phi)$.
Sometimes, these approaches provide compact expressions for the effective
viscosity, although usually a multipole expansion must be done in order to
obtain numerical values \cite{fixman,beenakker,saarloos,bedeaux}.

Beenakker \cite{beenakker}, by neglecting Brownian motion, gave a theory for
the concentration and wave-vector dependence of the viscosity valid up to high
concentrations that fully takes into account the many-body hydrodynamic
interactions between an arbitrary number of spheres. Batchelor and Green
\cite{batchelor} developed a theory for the evaluation of the bulk stress in a
semi-dilute suspensions of hard spheres. By neglecting Brownian motion and
assuming a random particle distribution they found the following expression
for the viscosity
\begin{equation}
\eta\left(  \phi\right)  =\eta_{0}\left(  1+\frac{5}{2}\phi+5.2\phi
^{2}\right)  ,\label{batchelor1}%
\end{equation}
where $\eta_{0}$ is the solvent viscosity and $\phi$ is the volume fraction.
This result has been regarded as the high-frequency value of the real part of
the complex shear viscosity \cite{vanderwerff1}. A different result is
obtained if Brownian motion is taken into account. Then appears an extra term
accounting for the stresses generated in the dispersion by the random
movements of the particles. Numerical calculations performed in Ref.
\cite{batchelor2} lead to the formula
\begin{equation}
\eta\left(  \phi\right)  =\eta_{0}\left(  1+\frac{5}{2}\phi+6.17\phi
^{2}\right)  ,\label{batchelor2}%
\end{equation}
considered as the low-shear and zero-frequency value of the real part of the
complex shear viscosity because in the calculation Brownian motion dominates
\cite{vanderwerff1}.

Bedeaux and collaborators \cite{bedeaux} introduced a general formalism for
the calculation of the effective shear viscosity for a suspension of spheres
as a function of the wave vector and the frequency for arbitrary values of the
volume fraction. Using the conservation law for the total momentum density of
the fluid and the suspended spheres, they perform a multipole expansion of the
one sphere response function to second order in the wave-vector and an
expansion in the volume fraction up to the second order, obtaining the
expression
\begin{equation}
\eta\left(  \phi\right)  =\eta_{0}\left(  1+\frac{5}{2}\phi+4.8\phi
^{2}\right)  .\label{bedeaux}%
\end{equation}
Russel and Gast \cite{russel} formulated a theory for the nonequilibrium
structure and stresses in a sheared suspension with a fluid rest state. Many
body interactions were handled exactly in the thermodynamics but truncated at
the pair level for the hydrodynamics.

More recently, on the basis of the pair Smoluchowski equation, Cichocki and
Felderhof \cite{cichocki1} have obtained exact results for $\eta\left(
\phi,\omega\right)  $ to $O(\phi^{2})$. Without Brownian motion contributions,
their result for $\omega=0$ is
\begin{equation}
\eta\left(  \phi\right)  =\eta_{0}\left(  1+\frac{5}{2}\phi+5.00\phi
^{2}\right)  ,\label{cichocki1}%
\end{equation}
while with Brownian motion contributions they find
\begin{equation}
\eta\left(  \phi\right)  =\eta_{0}\left(  1+\frac{5}{2}\phi+5.91\phi
^{2}\right)  .\label{cichocki2}%
\end{equation}
Verberg and collaborators \cite{verberg} presented an approximate theory based
on the contribution of two physically relevant processes taking place at
different time scales: (i) at short time scales the viscosity of the
suspension increases when compared to that of the pure solvent at infinite
dilution due to the finite probability to find two particles at contact,
whereas (ii) at long times it is difficult for a Brownian particle to diffuse
out of the cage formed around it by its neighbors. For $\omega=0$ these
authors obtain the approximate result%

\begin{equation}
\eta\left(  \phi\right)  =\eta_{0}\left(  1+\frac{5}{2}\phi+6.03\phi
^{2}\right)  ,\label{ververg1}%
\end{equation}
while for $\omega\rightarrow\infty$ they propose
\begin{equation}
\eta\left(  \phi\right)  =\eta_{0}\frac{1-0.5\phi}{\left(  1-\phi\right)
^{3}},\label{ververg2}%
\end{equation}
which is accurate for all $0<\phi<0.55$. Although their results compare
reasonably well with experimental data, they are based in a number of
assumptions not completely justified, as mentioned in Ref. \cite{verberg}.

The simple formulas reviewed are of fundamental importance for the
understanding of the rheology of suspensions. However, these formulas are far
from satisfactory since they only lead to good quantitative results in the low
volume fraction regime \cite{mellema,vanderwerff1,vanderwerff2}.

This restriction have lead to many works searching for simple procedures or
semi-empirical expressions to calculate the viscosity of suspensions and
dispersions at frequencies and shear rates relevant to industrial applications
\cite{dames}. Several phenomenological formulas\cite{krieger,quemada} have
been proposed in order to fit experiments in the largest possible range of
volume fractions. For example, Krieger and Dougherty found empirically
\cite{krieger} that the expression
\begin{equation}
\eta\left(  \phi\right)  =\eta_{0}\left(  1-\frac{\phi}{\phi_{c}}\right)
^{-2.5\phi_{c}},\label{krieger}%
\end{equation}
with $\phi_{c}$ the filling fraction at maximum packing, agrees reasonably
well with the experimental data. Moreover, this relation reduces to the
correct Einstein equation in the limit of infinite dilution. Other example is
the following expression obtained by Quemada \cite{quemada}
\begin{equation}
\eta\left(  \phi\right)  =\eta_{0}\left(  1-\frac{\phi}{\phi_{c}}\right)
^{-2},\label{quemada}%
\end{equation}
which gives the right divergence for $\eta$ as $\phi$ approaches $\phi_{c} $,
but, it does not reduces to Einstein's expression for low $\phi$. Nonetheless,
although useful, these expressions lack of physical support and therefore do
not provide insight about the phenomenology of the system.

In this article, we explore a different approach to the problem leading to a
simple an quantitatively powerful description of concentrated suspensions of
mono- and poly-disperse suspensions of colloidal particles. Our model assumes
that at macroscopic length and time scales, the suspension can be considered
as a truly continuum medium. However, the precise formulas are derived only by
considering the discrete nature of the suspended phase through its possible
configurations and poly-dispersion.

The technique we will adopt has been introduced long time ago in the context
of different physical properties of dispersed systems\cite{larson}%
-\cite{vandeven}, including optical\cite{bruggeman,fuchs,chang} and nonlocal
dielectric systems \cite{mendoza}. In the case of rheological properties, a
limited success has been obtained when comparing with experimental results
because the geometrical information of the system has poorly been included
\cite{larson,vandeven,arrhenius,ball}. In this work, we improve previous
results by incorporating structural information from the start through an
effective filling fraction. Moreover, in order to better understand some of
the assumptions implicit in the differential technique, we perform a
mesoscopic analysis of the problem by using Fokker-Planck equations
\cite{ISHPRE,brady04,drossinos}. This analysis allows us to elucidate the main
assumptions and approximations underlying the differential method.

The article is organized as follows, in Section II we introduce our
differential effective medium model. In Section III we perform an analysis of
the model based on a mesoscopic and hydrodynamic approach. In Section IV we
compare the predictions of our model with previous theories and with various
experimental results. Finally Section IV is devoted to conclusions.

\section{Differential viscosity model}

At very low volume fractions, the shear viscosity $\eta$ of a suspension of
hard spheres can be predicted by the simple formula \cite{einstein}
\begin{equation}
\eta\left(  \phi\right)  =\eta_{0}\left(  1+\frac{5}{2}\phi\right)
,\label{einstein}%
\end{equation}
that Einstein calculated from the viscous dissipation produced by the flow
around a single sphere. Notice that expression (\ref{einstein}) does not take
into consideration the way in which the spheres are arranged in the sample.
Therefore it contains no information about the maximum packing of spheres the
system may allocate. This information is not relevant for very low volume
fractions, nonetheless it does matter for higher concentration of spheres.
Even more, Einstein's expression contains no information about the radii of
the spheres and it may be applied to any polydisperse distribution of spheres
at low volume fractions giving to it its universal character.

In order to extend Einstein's expression to larger filling fractions $\phi$ we
use a differential effective medium approach. For this purpose, it is
important to consider that experimental results show that the viscosity
diverges at volume fractions $\phi<1$. This is related to the fact that the
spheres can not occupy all the volume of the sample due to geometrical
restrictions. For example, for a face centered cubic (FCC) arrangement of
identical spheres, the maximum volume that the spheres may occupy is larger
than for a random arrangement of spheres. At some point, the spheres become
arrested giving rise to the divergence of the viscosity. These considerations
suggest that the relevant quantity in expression (\ref{einstein}) is not
$\phi$ but an \emph{effective} filling fraction $\phi_{eff}$ that incorporates
geometrical information. We propose the following expression for $\phi_{eff}$:%
\begin{equation}
\phi_{eff}=\frac{\phi}{1-c\phi},\label{phieff}%
\end{equation}
where $c$ is a structural or crowding factor that takes into account the
arrangements of the spheres in the suspension and is given by%

\begin{equation}
c=\frac{1-\phi_{c}}{\phi_{c}},\label{c}%
\end{equation}
with $\phi_{c}$ the filling fraction at maximum packing. The main
characteristic of $\phi_{eff}$ is that it is the most simple way to
incorporate geometrical information having the characteristic that $\phi
_{eff}\sim\phi$ at low volume fractions and $\phi_{eff}=1$ at maximum packing.
The introduction of $\phi_{eff}$ is analogous to the excluded volume
correction proposed by van der Waals to incorporate the finite size of the
molecules in the equation of state of the ideal gas. In fact, considering the
excluded volume correction, the equation of state of an ideal gas is replaced
by $p=\rho_{eff}k_{B}T$, with $\rho_{eff}=\rho/(1-v\rho)$. Here $\rho=N/V$ is
the number density of molecules in the volume $V$ and $v$ is the volume of
each molecule. In our case, the excluded volume is due not only to the finite
size of the spheres but also to the volume that is not accessible to other
spheres due to their geometrical arrangement.

The differential effective medium theory (DEMT) is based in a progressive
addition of spheres to the sample. Consider a suspension of spheres with a
filling fraction $\phi_{eff}$. Then, suppose that we increase by $\delta
\phi_{eff}$ the particle concentration in the suspension of viscosity
$\eta\left(  \phi_{eff}\right)  $ by adding a small quantity $\Delta\phi
_{eff}$ of few new particles. If we treat the suspension into which we add
these particles as a \emph{homogeneous} effective medium of viscosity
$\eta\left(  \phi_{eff}\right)  $, then the new viscosity can be written as%
\begin{equation}
\eta\left(  \phi_{eff}+\delta\phi_{eff}\right)  =\eta\left(  \phi
_{eff}\right)  \left(  1+\frac{5}{2}\Delta\phi_{eff}\right)
,\label{recursive1}%
\end{equation}
where the added quantity of new spheres is related to $\delta\phi_{eff}$ by%
\begin{equation}
\Delta\phi_{eff}=\frac{\delta\phi_{eff}}{1-\phi_{eff}}.\label{deltaphi}%
\end{equation}
The normalizing factor that appears in Eq.(\ref{deltaphi}) is due to the fact
that one has to remove part of the effective medium in order to allocate the
new particles, in other words, the new fraction of spheres is given by
$\phi_{eff}+\delta\phi_{eff}=\phi_{eff}(1-\Delta\phi_{eff})+\Delta\phi_{eff}$,
from which we find $\Delta\phi_{eff}$. Integrating Eq.(\ref{recursive1}) we
obtain%
\begin{equation}
\eta\left(  \phi\right)  =\eta_{0}\left(  1-\phi_{eff}\right)  ^{-5/2}%
.\label{viscosity1}%
\end{equation}
This expression is similar to the empirical proposals of \cite{krieger},
Eq.(\ref{krieger}), and \cite{quemada}, Eq.(\ref{quemada}), and also to other
differential effective medium theories\cite{pal2}-\cite{vandeven}. However,
our model crucially makes use of the geometrical information of the system
from the start through $\phi_{eff}$. This results in an huge improvement of
the quantitative description when compared to experimental data as will be
shown in Section IV. Thus, Eq. (\ref{viscosity1}) is a powerful expression
accounting for a variety of situations ranging from disordered to ordered
system of spheres, or to systems with poly-dispersion of sizes. Even more, by
an appropriate choice of the coefficient of the $\phi$ term in
Eq.(\ref{einstein}) it can also be applied to other systems ranging from
emulsions to arbitrarily shaped particles.

\section{Mesoscopic hydrodynamics analysis of the model}

The model of a concentrated suspension of particles formulated in the previous
section essentially considers the superposition of dilute suspensions of
particles added in a heat bath at different stages. At each stage, the heat
bath is considered as an effective continuum medium with an effective
viscosity depending on the volume fraction of the particles added up to the
previous stage. The assumption that the heat bath is a continuum medium has
physical implications that can be elucidated by performing a mesoscopic
analysis of the dynamics of the system.

In non-equilibrium conditions imposed by a stationary flow $\mathbf{v}%
_{0}(\mathbf{r})$, the dynamics of a dilute \textquotedblleft
gas\textquotedblright\ of Brownian particles of mass $m$ suspended in a
Newtonian heat bath can be analyzed by means of a Fokker-Planck equation for
the single-particle probability distribution function
\cite{drossinos,ISHPRE,brady04,ryskin}. When the volume fraction of the
suspended particles is finite, hydrodynamic interactions play an important
role and then the description is usually performed by using a $N-$particle
distribution function obeying a multivariate Fokker-Planck or Smoluchowski
equations \cite{lionberger,brady04,verberg}.

However, to understand the physical assumptions underlying the differential
model previously introduced, we will follow a similar procedure in which, at
each stage, the new particles added constitute a dilute suspension in an
effective continuum medium. Under this assumption, the dynamics of the system
can be performed by means of the single-particle probability distribution
$f(\mathbf{r},\mathbf{u},t)$, depending on time $t$ and the instantaneous
position $\mathbf{r}$ and velocity $\mathbf{u}$ of the particles. Assuming
that the corresponding fluctuation-dissipation theorem is valid for the actual
non-equilibrium steady-state \cite{drossinos,brady04}, it is possible to show
that $f$ obeys the Fokker-Planck equation
\begin{equation}
\frac{\partial}{\partial t}P+\nabla\cdot\left(  \mathbf{u}P\right)
=\beta\frac{\partial}{\partial\mathbf{u}}\cdot\left[  (\mathbf{u}%
-\mathbf{v}_{0})P+\frac{k_{B}T}{m} \frac{\partial P}{\partial\mathbf{u}%
}\right]  ,\label{F-P}%
\end{equation}
where $\nabla= \partial/\partial\mathbf{r}$, $k_{B}T$ is the thermal energy
(assumed constant) and $\beta$ is the friction coefficient per mass unit of a
particle. Here, $\mathbf{v}_{0}$ is the stationary solution of the
Navier-Stokes equation $\rho_{0}\mathbf{v}_{0}\nabla\mathbf{v}_{0}=\nabla
\cdot\!\!\,\,\mathbf{P}^{0}\!\!\,\,$ for the heat bath, with the appropriate
boundary conditions on the surface of the particle and $\rho_{0} $ is the
constant density of the bath. Equations similar to (\ref{F-P}) have been
derived for related systems where the external flow introduces corrections
leading to the breaking of the fluctuation-dissipation theorem
\cite{ryskin,ISHPRE,rosalio}. Here, we will assume that Eq. (\ref{F-P}) is
valid for arbitrary externally applied velocity gradients.

Eq. (\ref{F-P}) allows one to calculate a set of evolution equations for the
moments of the probability distribution, constituting the \textquotedblleft
hydrodynamic\textquotedblright\ description of the system
\cite{ISHPRE,materiales}. Taking into account the usual definitions for the
hydrodynamic-like fields: mass density $\rho(\mathbf{r},t)$, velocity
$\mathbf{v}(\mathbf{r},t)$ and pressure tensor $\!\!\,\mathbf{P}%
^{k}\!\!\,(\mathbf{v},t)$ (see reference \cite{ISHPRE}), the evolution
equation for $\rho$ is found to be
\begin{equation}
\frac{\partial\rho}{\partial t}=-\nabla\cdot\mathbf{J}_{D},\label{contmasa}%
\end{equation}
where the diffusion current $\mathbf{J}_{D}(\mathbf{r},t)=\rho\mathbf{v}$
obeys \cite{SteadyState}
\begin{equation}
\mathbf{J}_{D}+\beta^{-1} \frac{D\mathbf{J}_{D}}{Dt}=\rho\mathbf{v}_{0}
-\beta^{-1}\left[  \nabla\cdot\,\,\!\!\mathbf{P}^{k}\!\!\,\,\right]
.\label{diffusive}%
\end{equation}
Here $D/Dt$ is the material time derivative. Eq. (\ref{diffusive}) contains
the pressure tensor $\,\!\!\mathbf{P}^{k}\!\!\,$ of the suspended phase whose
evolution equation is given by \cite{ISHPRE,materiales}
\begin{equation}
2 \!\!\,\,\mathbf{P}^{k}\!\!\,\,+\beta^{-1}\frac{D}{Dt}\!\!\,\,\mathbf{P}%
^{k}\!\!\,\, =2\frac{k_{B}T}{m} \rho\mathbf{1}- 2 \rho D_{0} \nabla
\mathbf{v}^{s},\label{evolution-P}%
\end{equation}
where the symbol $s$ stands for the symmetric part of a tensor, $\mathbf{1}$
is the unit tensor and $D_{0}=\frac{k_{B}T}{m\beta}$ is the single-particle
diffusion coefficient. In writing the second term at the left-hand side of Eq.
(\ref{diffusive}), we have neglected contributions of the order $\beta^{-2}$.
The last equation can be separated into two equations by considering
$\!\!\,\mathbf{P}^{k}\,=p\mathbf{1}+\mathbf{L}$, with $\mathbf{L}(\vec{r},t)$
the traceless stress tensor. Then, one obtains an evolution equation for the
pressure $p$, and other one for the non-diagonal terms $\mathbf{L}$,
associated with the dissipation in the system. Assuming that the diagonal part
relaxes faster than the other components, one has the state equation
$p=\frac{k_{B}T}{m} \rho$. On the other hand, the resulting evolution equation
for $\mathbf{L}$ can be recasted in the form
\begin{equation}
2 \mathbf{L}+\beta^{-1}\frac{D}{Dt}\mathbf{L} =- 2 \rho D_{0} (\nabla
\mathbf{v}_{0})^{0}- 2 \rho D_{0} (\nabla\mathbf{J})^{0},\label{evolution-L}%
\end{equation}
where we have used the definition of the local diffusion current
$\mathbf{J}=\mathbf{J}_{D}-\rho\mathbf{v}_{0}=\rho(\mathbf{v}- \mathbf{v}%
_{0})$ and neglected terms of the order $\rho^{-1}\nabla\rho$.

An evolution equation for the total stress tensor $\!\!\,\,\mathbf{Q}%
^{0}\!\!\,\,$ of the suspension can be obtained from Eq. (\ref{evolution-L})
by considering the symmetric traceless part of the stress tensor of the heat
bath: $\!\!\,\,\mathbf{P}^{0}\,\,\!\!=2\eta_{0}\left(  \nabla\vec{v}%
_{0}\right)  ^{0}$, where $\eta_{0}$ is the viscosity of the (Newtonian) bath
\cite{bedeaux}. Thus, multiplying Eq. (\ref{evolution-L}) by $-\rho_{0}%
\tilde{\rho}^{-1}$ with $\tilde{\rho}$ the average density of the suspended
phase, adding $\mathbf{P}^{0}$ and neglecting the convective term in the total
time derivative, one arrives at the following evolution equation for the total
traceless stress tensor of the suspension $\!\!\,\,\mathbf{Q}^{0}%
\!\!\,\,\equiv\!\!\,\mathbf{\,P}^{0}\!\!\,\,-2\!\!\,\,\rho_{0}\tilde{\rho}
^{-1}\mathbf{L}\!\!\,\,$ ,
\begin{equation}
\!\!\,\,\mathbf{Q}^{0}\,\,\!\!+\beta^{-1} \frac{\partial}{\partial
t}\!\!\,\,\mathbf{Q}\,\,\!\!^{0}=2\eta_{e}(\nabla\vec{v}_{0})\,^{0} +
2D_{0}\rho_{0}\tilde{\rho}^{-1} (\nabla\vec{J})^{0},\label{evol-Q}%
\end{equation}
where we have defined the effective viscosity coefficient $\eta_{e}$ of the
whole suspension as
\begin{equation}
\eta_{e}=\eta_{0}\left(  1+D_{0} \rho_{0} \eta_{0}^{-1} \right)
,\label{visceffect}%
\end{equation}
which takes into account the effects of Brownian motion of the particles. In
deriving Eqs. (\ref{evol-Q}) and (\ref{visceffect}) we have assumed that terms
of the order $D_{0}\rho^{-2}\mathbf{J}\cdot\nabla\rho$ are negligible. This is
consistent with the approximation $\Delta\rho/\rho\ll1$ used to derive Eq.
(\ref{evolution-L}).

The dynamic viscosity of the suspension as a function of the frequency
$\omega$ and the wave vector $\mathbf{k}$, can be calculated by linearizing
the hydrodynamic equations (\ref{contmasa})-(\ref{evol-Q}). With this purpose,
we will assume that $\rho=\tilde{\rho}\,+\delta\rho$, $\mathbf{v}%
_{0}=\mathbf{v}_{s}+\delta\mathbf{u}$, $\mathbf{L}=\delta\mathbf{L}$ and
$\mathbf{Q}^{0}=\mathbf{Q}_{0}^{0}+\delta\mathbf{Q}^{0}$, with $\delta\rho$,
$\delta\mathbf{u}$ and $\delta\mathbf{Q}^{0}$ the deviations with respect to
the average values $\tilde{\rho}$, $\mathbf{v}_{s}$ and $\mathbf{Q}_{0}^{0}$.
Here, $\delta\mathbf{Q}^{0}$ contains the effect on the stresses induced by
the particles when locally perturb the velocity field of the bath:
$\delta\mathbf{Q}^{0}=2\eta_{0}\nabla\delta\mathbf{u}^{0}-2\rho_{0}\tilde
{\rho}^{-1}\delta\mathbf{L}$.

Then, at first order in the deviations, from Eqs. (\ref{contmasa}),
(\ref{diffusive}) and (\ref{evol-Q}), one may derive the following set of
fluctuating hydrodynamic equations \cite{materiales}
\begin{equation}
\frac{\partial}{\partial t}\delta\rho+\mathbf{v}_{s}\cdot\nabla\delta
\rho=-\nabla\cdot\left(  \delta\mathbf{J}\right) ,\label{deltarho}%
\end{equation}

\begin{equation}
\delta\mathbf{J}+\beta^{-1}\frac{\partial}{\partial t}\delta\mathbf{J}= -D_{0}
\nabla\delta\rho-\frac{1}{2}\beta^{-1} \tilde{\rho} \rho_{0}^{-1} \nabla
\cdot\delta\mathbf{Q}^{0} -2\beta^{-1}\nabla\cdot\delta\mathbf{L}
,\label{deltaj}%
\end{equation}

\begin{equation}
\delta\mathbf{Q}^{0}+\beta^{-1}\frac{\partial}{\partial t}\delta\mathbf{Q}%
^{0}=2\mathbf{\eta}_{e}\left(  \nabla\delta\mathbf{u}\,\right)  ^{0}
+2D_{0}\tilde{\rho}\rho_{0}^{-1}\left(  \nabla\delta\mathbf{J}\right)
^{0}.\label{deltaq}%
\end{equation}

Assuming that $\mathbf{v}_{s}$ is a linear flow, then the correction
$\delta\mathbf{u}$ due to the presence of the particle is proportional to the
constant tensor $\nabla\mathbf{v}_{s}$ \cite{landau}. Following Ref.
\cite{landau}, Eq. (\ref{deltaq}) can be written in the form
\begin{equation}
\delta\mathbf{Q}^{0}+\beta^{-1}\frac{\partial}{\partial t}\delta\mathbf{Q}%
^{0}=5\phi\mathbf{\eta}_{e}\left(  \nabla\mathbf{v}_{s}\,\right)  ^{0}%
+2D_{0}\tilde{\rho}\rho_{0}^{-1}\left(  \nabla\delta\mathbf{J}\,\right)
^{0}.\label{deltaqB}%
\end{equation}
An equation similar to (\ref{deltaqB}) can be derived for $\delta\mathbf{L}$.
Taking the space-time Fourier transform of Eqs. (\ref{deltarho}),
(\ref{deltaj}) and (\ref{deltaqB}), one may solve for $\delta\mathbf{Q}^{0}$.
Assuming for simplicity that $\mathbf{k}=\left(  k,0,0\right)  $ and
neglecting terms smaller than $\beta^{-2}$, one obtains
\begin{equation}
\delta{Q}_{xy}^{0}=\frac{5\phi\eta_{e}}{1-i\omega\beta^{-1}-\frac{D_{0}%
\beta^{-1}k^{2}}{\left(  1-i\omega\beta^{-1}\right)  }}(\nabla\mathbf{v}%
_{s})_{xy}.\label{dynamicviscosity1}%
\end{equation}
Using Eq. (\ref{visceffect}) and $\mathbf{Q}^{0}=\mathbf{Q}_{0}^{0}%
+\delta\mathbf{Q}^{0}$, we obtain a dynamic viscosity which follows a second
order continued-fraction expansion
\begin{equation}
\eta(\omega,k;\phi)\simeq\eta_{0}\left[  1+\frac{5}{2}\phi\frac{1+Sc^{-1}%
}{1-i\omega\beta^{-1}-\frac{D_{0}\beta^{-1}k^{2}}{\left(  1-i\omega\beta
^{-1}\right)  }}\right]  ,\label{viscfinal}%
\end{equation}
where we have introduced the Schmidt number $Sc=\eta_{0}(D_{0}\rho_{0})^{-1}$,
which for typical suspensions is usually a small quantity $<10^{-3}$. By
expanding Eq. (\ref{viscfinal}) up to first order in $\omega$ and second order
in $k$, we obtain an expression for the viscosity similar to that reported in
Ref. \cite{bedeaux}. This $(\omega,k)$-dependent correction could be
significative for very small times $t\sim\beta^{-1}$. Therefore, at
(diffusion) times $t\gg\beta^{-1}$, Eq. (\ref{viscfinal}) reduces to
Einstein's formula: $\eta(\phi)\simeq\eta_{0}\left[  1+\frac{5}{2}\phi\right]
$.

\subsection{Generalization to finite volume fractions}

To generalize results obtained in the previous section to the case of finite
volume fractions, we will follow arguments similar to those of Section
\textbf{II}, but applied to the long-time limit of the evolution equations
derived in the previous section. Then, the subindex $i$ will be used to
indicate the stage $i$ in which, according to Einstein's relation, the
effective viscosity is $\eta(\phi_{i})\simeq\eta_{0}\left[  1+\frac{5}{2}%
\phi_{i}\right] $, with the volume fraction at stage $i$, $\phi_{i}$, defined
as in Eq. (\ref{phieff}) .

Thus, assuming that the suspension of particles at stage $i$ plays the role of
an effective (continuum) medium, it is clear that the probability distribution
of the new phase ($i+1$) will satisfy a Fokker-Planck equation similar to Eq.
(\ref{F-P}), where now the friction coefficient can be denoted by $\beta
_{i+1}$ and the diffusion coefficient by $D_{0}^{(i+1)}=k_{B}T/(m\beta_{i+1}%
)$. Following a similar procedure as the already indicated in this section,
from the Fokker-Planck equation at stage $i+1$ the linearized set of evolution
equations for the fluctuations of the hydrodynamic fields is
\begin{equation}
\frac{\partial}{\partial t}\delta\rho_{i+1} +\mathbf{v}_{s}\cdot\nabla
\delta\rho_{i+1} =-\nabla\cdot\left(  \delta\mathbf{J}_{i+1}\right)
,\label{deltarho2}%
\end{equation}

\begin{equation}
\delta\mathbf{J}_{i+1}+\beta_{i+1} ^{-1}\frac{\partial}{\partial t}%
\delta\mathbf{J}_{i+1}= -D_{0}^{i+1} \nabla\delta\rho_{i+1} -\frac{1}{2}%
\beta_{i+1} ^{-1} \tilde{\rho}_{i+1} \rho_{0}^{-1} \nabla\cdot\delta
\mathbf{Q}^{0}_{i+1} -2\beta_{i+1} ^{-1}\nabla\cdot\delta\mathbf{L}_{i+1}
,\label{deltaj2}%
\end{equation}

\begin{equation}
\delta\mathbf{Q}^{0}_{i+1}+\beta_{i+1}^{-1}\frac{\partial}{\partial t}%
\delta\mathbf{Q}^{0}_{i+1}=2\mathbf{\eta}_{e,i}\left(  \nabla\delta
\mathbf{u}\,\right)  ^{0} +2D_{0}^{i+1}\tilde{\rho}_{i+1}\rho_{0}^{-1}\left(
\nabla\delta\mathbf{J}_{i+1}\right)  ^{0}.\label{deltaq2}%
\end{equation}
where we have used the definition $\mathbf{Q}^{0}_{i+1} \equiv\mathbf{Q}%
^{0}_{i} - 2 \rho_{0}\tilde{\rho}^{-1}_{i+1}\mathbf{L}_{i+1}$. In writing Eq.
(\ref{deltaq2}) an important assumption has been done: We have assumed that
the traceless stress tensor $\mathbf{Q}^{0}_{i}$ of the effective medium has
relaxed to its stationary value, thus implying that only the relaxation time
$\beta_{i+1}^{-1}$ enters into Eq. (\ref{deltaq2}). This hypothesis allows us
to obtain the following relation for the effective viscosity of the system at
the stage $i+1$:
\begin{equation}
\eta(\phi_{i+1})\simeq\eta_{i}(\phi_{i})\left[  1+\frac{5}{2}\phi_{i+1}\right]
.\label{viscStage2}%
\end{equation}

If we now take into account that the total volume fraction occupied by the
particles at the stage $i+1$ is given by $\phi=\phi_{i}(1-\phi_{i+1}%
)+\phi_{i+1}$ and assume that at each stage a very small (differential)
quantity of particles is added, then: $\phi_{i+1} = \delta\phi/(1-\phi_{i})$,
with $\delta\phi= \phi-\phi_{i}$. Substituting this relation into Eq.
(\ref{viscStage2}), we finally obtain the differential relation ${\delta\eta
}/{\eta_{i}}\simeq({5}/{2}){\delta\phi}/{(1-\phi_{i})}$ with $\delta\eta
=\eta_{i+1}-\eta_{i}$ that after integration leads to
\begin{equation}
\eta\left(  \phi\right)  =\eta_{0}\left(  1-\phi\right) ^{-5/2}%
,\label{viscStage4}%
\end{equation}
which has the same form of Eq. (\ref{viscosity1}). As in Section \textbf{II},
it must be emphasized that geometrical information must be introduced by
taking into account that Eq. (\ref{viscStage4}) depends upon the effective
volume fraction $\phi_{eff}$ defined in Eq. (\ref{phieff}).

\section{Results and comparison with experiments}

In this section we compare the predictions of the improved DEMT with several
experimental results and with the results obtained from other theories. To
carry out these comparisons, it is important to notice that the value of
$\phi_{c}$ in Eq.(\ref{c}) is a free parameter of the theory to be chosen in
order to best fit the experimental results. Nonetheless, this parameter can be
chosen beforehand based on physical arguments, and then used to compare with
specific experimental situations.

This is done in Fig. 1, where the behavior of the relative viscosity
$\eta\left(  \phi\right)  /\eta_{0}$ is compared with experimental results of
de Kruif \emph{et al.} \cite{de Kruif} and of Krieger \cite{krieger2} for high
and low-shear rates as a function of the volume fraction $\phi$. The lower
curve is the prediction of Eq.(\ref{viscosity1}) with $\phi_{c}=0.7404$, which
corresponds to FCC close packing. This gives an excellent agreement with the
experimental results for the case of high-shear rates. The upper curve
represents the prediction of Eq.(\ref{viscosity1}) with $\phi_{c}=0.63 $,
which corresponds to the random close packing (RCP) of identical spheres. The
comparison with the experimental results for low-shear rates is again
excellent. These results are consistent with the known fact that for low-shear
rates the spheres remain disordered while at high-shear rates the equilibrium
microstructure of the dispersion is completely destroyed and the spheres adopt
an ordered FCC configuration. In Fig. 2 we use again the value $\phi_{c}=0.63$
to compare our theory with the experimental results obtained by Saunders
\cite{saunders}, by Krieger and coworkers\cite{krieger2}, and by
Kops-Werkhoven and Fijnaut\cite{kops}. We also plot the calculated values
obtained by Beenakker\cite{beenakker} with a theory that fully takes into
account the many-body hydrodynamic interactions between an arbitrary number of
spheres. Notice again that the agreement with the experimental results is
excellent over the entire fluid range.

In Fig. 3 we compare $\eta\left(  \phi\right)  /\eta_{0}$ with the reduced
viscosity measurements by van der Werff \emph{et al.} \cite{vanderwerff1} and
by Zhu \emph{et al.} \cite{zhu} at high frequencies. We also plot the values
obtained by Cichocki and Felderhof \cite{cichocki2} and the predictions using
a proposal by Verberg \emph{et al.} \cite{verberg} which considers the
fraction of colloidal particle pairs at contact, Eq. (\ref{ververg2}). We used
as a parameter to fit the data $\phi_{c}=0.8678$. We note that Verberg's
predictions lie below the experimental points while the results of the
improved DEMT fits the data very well. Similarly, Fig. 4 shows the zero
frequency low-shear viscosity as obtained by our model using $\phi_{c}=0.63$,
and it is compared with experimental results and the model of Verberg \emph{et
al.} \cite{verberg}. Again, a very good agreement is found.

A more sensitive representation of the relative viscosity data is the one
proposed by Bedeaux \cite{bedeaux1}:
\begin{equation}
\frac{\eta\left(  \phi\right)  /\eta_{0}-1}{\eta\left(  \phi\right)  /\eta
_{0}+\frac{3}{2}}=\phi\left(  1+S\left(  \phi\right)  \right)  ,\label{S}%
\end{equation}
where $S(\phi)$ is an unknown function of the volume fraction. In Fig. 5, we
represent the function $S(\phi)$ for various experimental results obtained in
a variety of experimental situations and compare with the values obtained with
our model. We take the same values of $\phi_{c}$ used previously for the low-
and high-shear cases. The agreement with the experimental data is again very
good. The last set of experimental points corresponds to the high-frequency
limit of the real part of the complex viscosity as obtained by the linear
dynamic measurements performed by van der Werff and coworkers
\cite{vanderwerff1}. The best fit to the real part, obtained with $\phi
_{c}=0.8678$ gives an excellent agreement with the experimental values. As a
comparative, we also show the predictions for the high-frequency limit,
obtained by Beenakker \cite{beenakker}, by Batchelor and Green
\cite{batchelor}, by Russel and Gast \cite{russel}, and by Verberg \emph{et
al.} \cite{verberg}. In this very sensitive representation it is seen that
only the improved DEMT matches the data at all volume fractions.

Finally, let us notice that the functional form\ for the reduced viscosity in
the improved DEMT as given by Eq.(\ref{viscosity1}) makes no reference to any
material parameter neither to the geometrical information of the system. This
means that a plot of the reduced viscosity $\eta\left(  \phi\right)  /\eta
_{0}$ vs. $\phi_{eff}$ should be independent of the experimental details or of
the sample. This is confirmed in Fig. 6 where we plot $\left[  \eta\left(
\phi\right)  /\eta_{0}\right]  ^{-2/5}$ vs. $\phi_{eff}$ for all the data sets
considered in the previous figures. The straight line is the prediction of the
improved DEMT, Eq.(\ref{viscosity1}). As can be seen, the generality of the
correlation and the agreement with our model is quite remarkable.

Now we proceed to compare our expression for the viscosity with the functional
form of other theories. First, let us notice that putting $\phi=\phi_{c}$ in
the denominator of Eq.(\ref{phieff}) and replacing the factor $5/2$ by $2$ in
Eq.(\ref{einstein}) we obtain the well known empirical expression obtained by
Quemada \cite{quemada}, Eq.(\ref{quemada}). This expression although gives the
right divergence at high filling fractions does not have the correct low
volume fraction limit given by Einstein's expression.

On the other hand, putting $\phi_{eff}=\phi/\phi_{c}$ in the denominator of
Eq.(\ref{deltaphi}) we recover the empirical expression obtained by Krieger
and Dougherty \cite{krieger}, Eq.(\ref{krieger}). Nonetheless the virial
expansion of this expression gives values for the second virial coefficient,
$k_{2}$, that are too small \cite{vandeven}. A generalization of this model
but including a dependence of the crowding factor on the volume fraction was
proposed by van de Ven \cite{vandeven}. However this model has the shortcoming
that gives rise to an expression for the viscosity that is more involved and
it includes a second adjustable parameter besides $\phi_{c}$ which is chosen
in order to adjust the second virial coefficient to the values predicted by
other theories. In contrast, our proposal does not contain adjustable
parameters besides $\phi_{c}$ and the series expansion gives the right values
for the second virial coefficient in a natural way as it is shown below.

Expanding Eq.(\ref{viscosity1}) in power series of $\phi$ for specific choices
of $\phi_{c}$ allows one to compare the improved DEMT with other theories.
Choosing the RCP value for the volume fraction at maximum packing, $\phi
_{c}=0.63$, we obtain up to $O(\phi^{3})$
\[
\eta\left(  \phi\right)  /\eta_{0}=1+\frac{5}{2}\phi+5.84325\phi
^{2}+12.5673\phi^{3}+O(\phi^{4}),
\]
which compares very well up to second order with the result that Cichoki and
Felderhof \cite{cichocki2} have obtained on the basis of the pair Smoluchowski
equation which are exact $O(\phi^{2})$. With Brownian motion contributions
they find for the second virial coefficient $k_{2}=5.91$ [see
Eq.(\ref{cichocki2})]. Batchelor and Green \cite{batchelor} including Brownian
motion have obtained $k_{2}=6.17$, as shown in Eq.(\ref{batchelor2}). Our
result differs form the exact one by $1.2\%$, while the result by Batchelor
and Green differs by $4.4\%$.

Taking FCC close packing $\phi_{c}=0.7404$, and expanding Eq.(\ref{viscosity1}%
) up to $O(\phi^{3})$ we obtain
\[
\eta\left(  \phi\right)  /\eta_{0}=1+\frac{5}{2}\phi+5.25155\phi
^{2}+9.93777\phi^{3}+O(\phi^{4}),
\]
which compares well up to second order with the result by Bedeaux and
collaborators \cite{bedeaux}, $k_{2}=4.8$, as shown in Eq.(\ref{bedeaux}) and
even better with the result of Cichocki and Felderhof without including
Brownian motion $k_{2}=5.00$, as shown in Eq.(\ref{cichocki1}). Our result
differs form the exact one by $5\%$, while the result by Bedeaux by $4\%$.

Using the value that best fits the high-frequency regime $\phi_{c}=0.8678$,
one finds for the real part of the complex relative viscosity
\[
\eta\left(  \phi\right)  /\eta_{0}=1+\frac{5}{2}\phi+4.75566\phi
^{2}+7.95275\phi^{3}+O(\phi^{4}),
\]
which compares well with the result by Batchelor and Green \cite{batchelor},
$k_{2}=5.2$, as shown in Eq.(\ref{batchelor1}) and in which Brownian motion is
not taken into consideration.

Finally, it is interesting to find a value for $\phi_{c}$ that best fits the
theoretical results obtained by Verberg \emph{et al.} \cite{verberg}. This
value is $\phi_{c}=0.95238$ and produces the following series expansion up to
$O(\phi^{7})$, just to show a few terms%

\[
\eta\left(  \phi\right)  /\eta_{0}=1+\frac{5}{2}\phi+4.5\phi^{2}+7.006\phi
^{3}+10.041\phi^{4}+13.6358\phi^{5}+17.8297\phi^{6}+22.6673\phi^{7}+O(\phi
^{8}),
\]
which is exact up to $O(\phi^{2})$ with the series expansion of the result by
Verberg \emph{et al.}which reads%

\[
\eta\left(  \phi\right)  /\eta_{0}=1+\frac{5}{2}\phi+4.5\phi^{2}+7\phi
^{3}+10\phi^{4}+13.5\phi^{5}+17.5\phi^{6}+22\phi^{7}+O(\phi^{8}),
\]
and the higher order terms are also amazingly close.

\section{Conclusions}

In summary, we have formulated an improved differential effective medium
theory that accounts for the dependence of the viscosity of hard-spheres
suspensions at arbitrary volume fractions. This theory takes into
consideration geometrical information of the system from the start through an
effective filling fraction and assumes a hierarchy of relaxation times, as we
have shown by performing a mesoscopic analysis based on Fokker-Planck equations.

The agreement with experimental data at low- and high-shear rates and in the
high-frequency limit is remarkable and superior to other theories. Also, the
virial expansion of our model compares well with the second virial coefficient
obtained by a number of different theories in all the regimes considered. In
the particular case of the high-frequency limit, it is possible to find a
value for $\phi_{c}$ for which the virial expansion is amazingly close to
higher orders to that of the model by Verberg \emph{et al}. Although our
proposal can be generalized to describe the viscoelastic behavior at
$\omega\neq0$ by simply allowing $\phi_{c}$ to take complex values, we have
not a derivation based on simple physical arguments to justify this
generalization. Such a goal requires further research.

\qquad{\LARGE Acknowledgements}

This work was supported in part by Grants DGAPA IN-107607 (CIM) and IN-108006 (ISH).

\newpage%

%TCIMACRO{\FRAME{ftbpFU}{3.9885in}{3.442in}{0pt}{\Qcb{Relative viscosity
%$\eta\left(  \phi\right)  /\eta_{0}$ at low and high shear rates as a function
%of the volume fraction $\phi$. The lines correspond to the\ predictions of the
%improved DEMT with RCP $\phi_{c}=0.63$ (upper line) and close packing at FCC
%$\phi_{c}=0.7404$ (lower line). The measured data are from Refs. \cite{de
%Kruif} (circles) and \cite{krieger2} (triangles).}}{}{fig1arxiv.eps}%
%{\special{ language "Scientific Word";  type "GRAPHIC";
%maintain-aspect-ratio TRUE;  display "USEDEF";  valid_file "F";
%width 3.9885in;  height 3.442in;  depth 0pt;  original-width 3.9401in;
%original-height 3.3961in;  cropleft "0";  croptop "1";  cropright "1";
%cropbottom "0";  filename 'Figures/fig1arxiv.eps';file-properties "XNPEU";}}}%
%BeginExpansion
\begin{figure}
[ptb]
\begin{center}
\includegraphics[
height=3.442in,
width=3.9885in
]%
{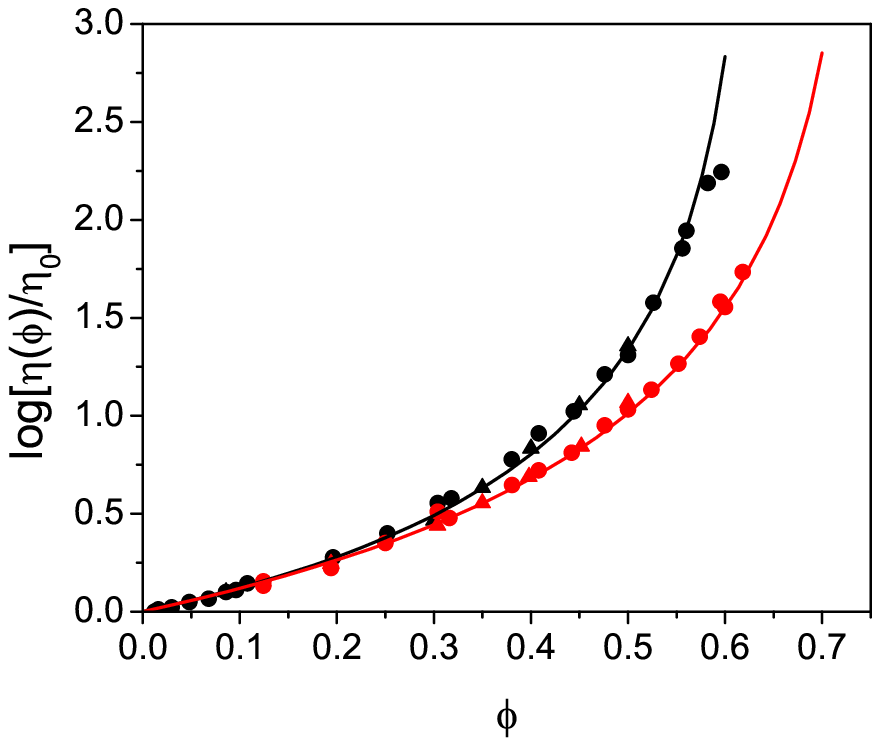}%
\caption{Relative viscosity $\eta\left(  \phi\right)  /\eta_{0}$ at low and
high shear rates as a function of the volume fraction $\phi$. The lines
correspond to the\ predictions of the improved DEMT with RCP $\phi_{c}=0.63$
(upper line) and close packing at FCC $\phi_{c}=0.7404$ (lower line). The
measured data are from Refs. \cite{de Kruif} (circles) and \cite{krieger2}
(triangles).}%
\end{center}
\end{figure}
%EndExpansion
%

%TCIMACRO{\FRAME{ftbpFU}{4.088in}{3.442in}{0pt}{\Qcb{Volume fraction dependence
%of the reciprocal of the relative viscosity $\eta_{0}/\eta\left(  \phi\right)
%$. The solid line is the result of the improved DEMT with $\phi_{c}=0.63$. The
%dashed line is taken from Ref. \cite{beenakker}. The measured data are from
%refs. \cite{saunders} (squares), \cite{krieger2} (triangles), and \cite{kops}
%(circles).}}{}{fig2arxiv.eps}{\special{ language "Scientific Word";
%type "GRAPHIC";  maintain-aspect-ratio TRUE;  display "USEDEF";
%valid_file "F";  width 4.088in;  height 3.442in;  depth 0pt;
%original-width 4.0369in;  original-height 3.3961in;  cropleft "0";
%croptop "1";  cropright "1";  cropbottom "0";
%filename 'Figures/fig2arxiv.eps';file-properties "XNPEU";}}}%
%BeginExpansion
\begin{figure}
[ptb]
\begin{center}
\includegraphics[
height=3.442in,
width=4.088in
]%
{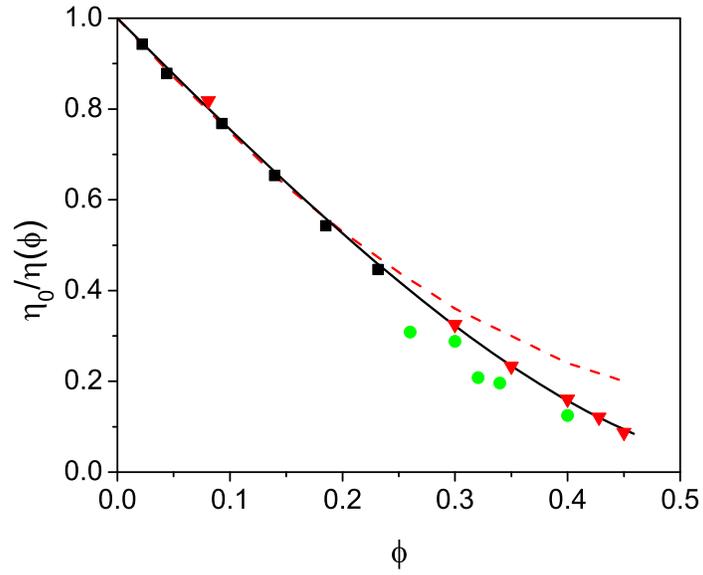}%
\caption{Volume fraction dependence of the reciprocal of the relative
viscosity $\eta_{0}/\eta\left(  \phi\right)  $. The solid line is the result
of the improved DEMT with $\phi_{c}=0.63$. The dashed line is taken from Ref.
\cite{beenakker}. The measured data are from refs. \cite{saunders} (squares),
\cite{krieger2} (triangles), and \cite{kops} (circles).}%
\end{center}
\end{figure}
%EndExpansion
%

%TCIMACRO{\FRAME{ftbpFU}{4.0465in}{3.3434in}{0pt}{\Qcb{Relative
%infinite-frequency viscosity $\eta\left(  \phi\right)  /\eta_{0}$ as a
%function of the volume fraction $\phi$. Squares, Zhu \QTR{em}{et al.}
%\cite{zhu}; circles, van der Werff \QTR{em}{et al.} \cite{vanderwerff1};
%triangles, Cichocki and Felderhof \cite{cichocki2}. The dashed line
%corresponds to the prediction by Verberg \QTR{em}{et al.} \cite{verberg}, and
%the solid line is the result of the improved DEMT with $\phi_{c}=0.8678$.}}%
%{}{fig3arxiv.eps}{\special{ language "Scientific Word";  type "GRAPHIC";
%maintain-aspect-ratio TRUE;  display "USEDEF";  valid_file "F";
%width 4.0465in;  height 3.3434in;  depth 0pt;  original-width 3.9972in;
%original-height 3.2984in;  cropleft "0";  croptop "1";  cropright "1";
%cropbottom "0";  filename 'Figures/fig3arxiv.eps';file-properties "XNPEU";}}}%
%BeginExpansion
\begin{figure}
[ptb]
\begin{center}
\includegraphics[
height=3.3434in,
width=4.0465in
]%
{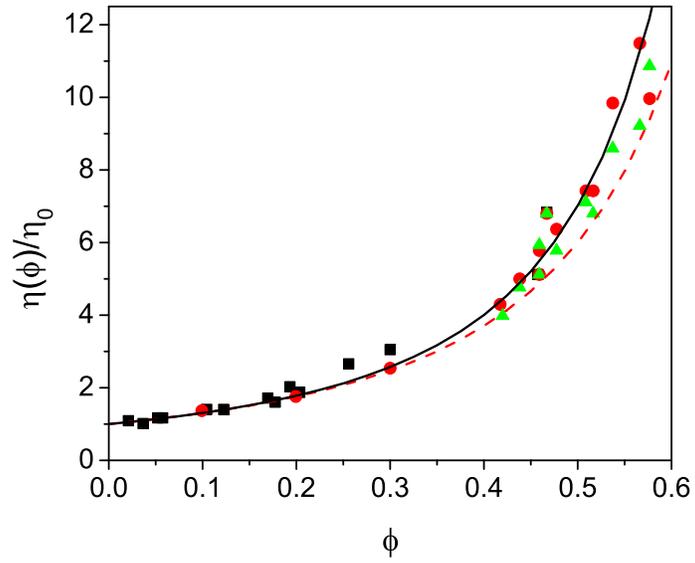}%
\caption{Relative infinite-frequency viscosity $\eta\left(  \phi\right)
/\eta_{0}$ as a function of the volume fraction $\phi$. Squares, Zhu \emph{et
al.} \cite{zhu}; circles, van der Werff \emph{et al.} \cite{vanderwerff1};
triangles, Cichocki and Felderhof \cite{cichocki2}. The dashed line
corresponds to the prediction by Verberg \emph{et al.} \cite{verberg}, and the
solid line is the result of the improved DEMT with $\phi_{c}=0.8678$.}%
\end{center}
\end{figure}
%EndExpansion
%

%TCIMACRO{\FRAME{ftbpFU}{4.0465in}{3.442in}{0pt}{\Qcb{Relative zero-frequency
%viscosity $\eta\left(  \phi\right)  /\eta_{0}$ at low shear rate as a function
%of the volume fraction $\phi$. Triangles, van der Werff \QTR{em}{et al.}
%\cite{vanderwerff1}; crosses, van der Werff and de Kruif \cite{vanderwerff2};
%circles, Jones \QTR{em}{et al. }\cite{jones}; squares, Papir and Krieger
%\cite{papir}. The dashed line corresponds to the prediction by Verberg
%\QTR{em}{et al.} \cite{verberg} and the solid line is the result of the
%improved DEMT with $\phi_{c}=0.63$.}}{}{fig4arxiv.eps}%
%{\special{ language "Scientific Word";  type "GRAPHIC";
%maintain-aspect-ratio TRUE;  display "USEDEF";  valid_file "F";
%width 4.0465in;  height 3.442in;  depth 0pt;  original-width 3.9972in;
%original-height 3.3961in;  cropleft "0";  croptop "1";  cropright "1";
%cropbottom "0";  filename 'Figures/fig4arxiv.eps';file-properties "XNPEU";}}}%
%BeginExpansion
\begin{figure}
[ptb]
\begin{center}
\includegraphics[
height=3.442in,
width=4.0465in
]%
{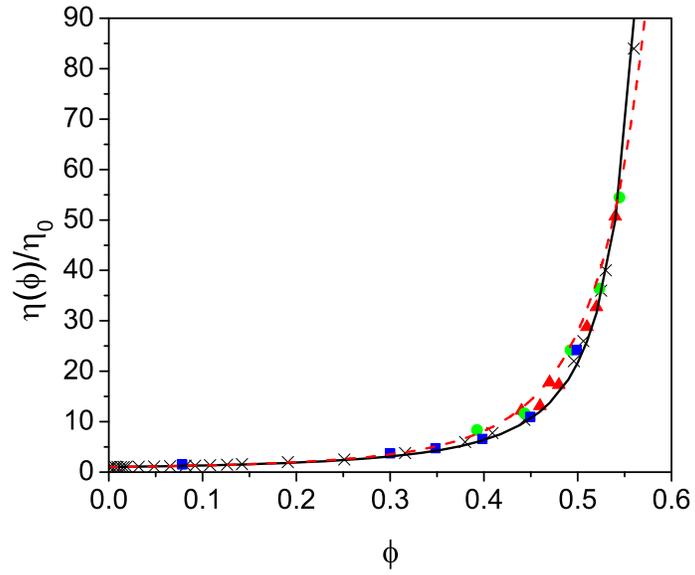}%
\caption{Relative zero-frequency viscosity $\eta\left(  \phi\right)  /\eta
_{0}$ at low shear rate as a function of the volume fraction $\phi$.
Triangles, van der Werff \emph{et al.} \cite{vanderwerff1}; crosses, van der
Werff and de Kruif \cite{vanderwerff2}; circles, Jones \emph{et al.
}\cite{jones}; squares, Papir and Krieger \cite{papir}. The dashed line
corresponds to the prediction by Verberg \emph{et al.} \cite{verberg} and the
solid line is the result of the improved DEMT with $\phi_{c}=0.63$.}%
\end{center}
\end{figure}
%EndExpansion
%

%TCIMACRO{\FRAME{ftbpFU}{3.9764in}{3.3434in}{0pt}{\Qcb{Representation of
%various viscosity data as suggested by Bedeaux (Ref. \cite{bedeaux1}).
%Squares, SJ18 low-shear limit (Refs. \cite{einstein} and \cite{de Kruif});
%circles, SJ18 high-shear limit (Refs. \cite{vanderwerff2} and \cite{de
%Kruif}); triangles, high-frequency limit of the real part of the complex shear
%viscosity (Ref.\cite{vanderwerff1}). Thick solid lines are the results of the
%improved DEMT with $\phi_{c}=0.63$ (upper), $\phi_{c}=0.7404$ (medium), and
%$\phi_{c}=0.8678$ (lower); thin dashed line is Beenakker's result (Ref.
%\cite{beenakker}), thin dotted line is Verberg \QTR{em}{et al}. prediction
%(Ref. \cite{verberg}), thin dash-dot line is the result of Russel and Gast
%(Ref. \cite{russel}), and the thin dash-dot-dot line is the prediction of
%Batchelor and Green (Ref. \cite{batchelor}).}}{}{fig5arxiv.eps}%
%{\special{ language "Scientific Word";  type "GRAPHIC";
%maintain-aspect-ratio TRUE;  display "USEDEF";  valid_file "F";
%width 3.9764in;  height 3.3434in;  depth 0pt;  original-width 3.9271in;
%original-height 3.2984in;  cropleft "0";  croptop "1";  cropright "1";
%cropbottom "0";  filename 'Figures/fig5arxiv.eps';file-properties "XNPEU";}}}%
%BeginExpansion
\begin{figure}
[ptb]
\begin{center}
\includegraphics[
height=3.3434in,
width=3.9764in
]%
{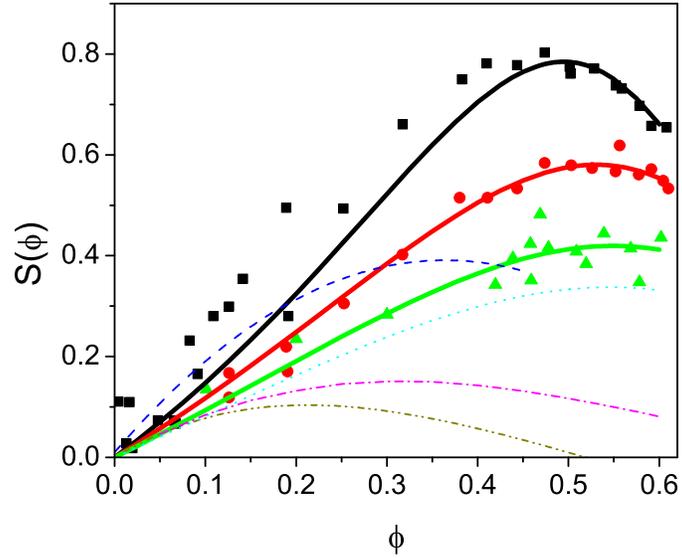}%
\caption{Representation of various viscosity data as suggested by Bedeaux
(Ref. \cite{bedeaux1}). Squares, SJ18 low-shear limit (Refs. \cite{einstein}
and \cite{de Kruif}); circles, SJ18 high-shear limit (Refs.
\cite{vanderwerff2} and \cite{de Kruif}); triangles, high-frequency limit of
the real part of the complex shear viscosity (Ref.\cite{vanderwerff1}). Thick
solid lines are the results of the improved DEMT with $\phi_{c}=0.63$ (upper),
$\phi_{c}=0.7404$ (medium), and $\phi_{c}=0.8678$ (lower); thin dashed line is
Beenakker's result (Ref. \cite{beenakker}), thin dotted line is Verberg
\emph{et al}. prediction (Ref. \cite{verberg}), thin dash-dot line is the
result of Russel and Gast (Ref. \cite{russel}), and the thin dash-dot-dot line
is the prediction of Batchelor and Green (Ref. \cite{batchelor}).}%
\end{center}
\end{figure}
%EndExpansion
%

%TCIMACRO{\FRAME{ftbpFU}{4.1303in}{3.4973in}{0pt}{\Qcb{Relative viscosity
%$\left[  \eta\left(  \phi\right)  /\eta_{0}\right]  ^{-2/5}$ as a function of
%the effective volume fraction $\phi_{eff}$. Symbols correspond to all the data
%sets of the previous figures and the solid line corresponds to the prediction
%of the improved DEMT, Eq.(\ref{viscosity1}).}}{}{fig6arxiv.eps}%
%{\special{ language "Scientific Word";  type "GRAPHIC";
%maintain-aspect-ratio TRUE;  display "USEDEF";  valid_file "F";
%width 4.1303in;  height 3.4973in;  depth 0pt;  original-width 4.0802in;
%original-height 3.4515in;  cropleft "0";  croptop "1";  cropright "1";
%cropbottom "0";  filename 'Figures/fig6arxiv.eps';file-properties "XNPEU";}}}%
%BeginExpansion
\begin{figure}
[ptb]
\begin{center}
\includegraphics[
height=3.4973in,
width=4.1303in
]%
{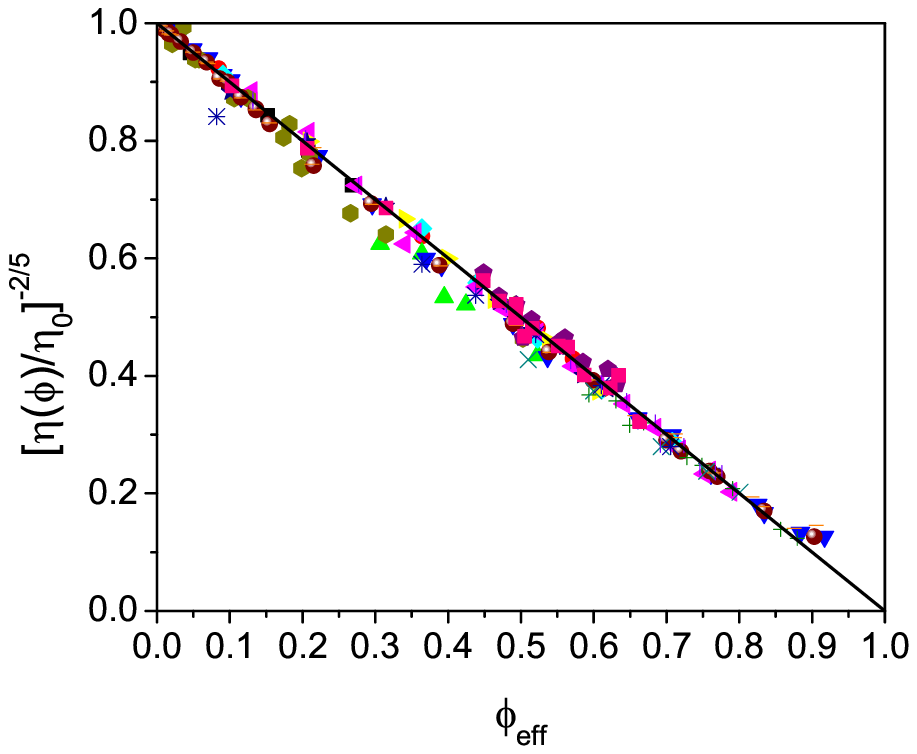}%
\caption{Relative viscosity $\left[  \eta\left(  \phi\right)  /\eta
_{0}\right]  ^{-2/5}$ as a function of the effective volume fraction
$\phi_{eff}$. Symbols correspond to all the data sets of the previous figures
and the solid line corresponds to the prediction of the improved DEMT,
Eq.(\ref{viscosity1}).}%
\end{center}
\end{figure}
%EndExpansion


\begin{references}
\bibitem{einstein} A. Einstein, \emph{Investigations on the Theory of Brownian Movement} (Dover, New York, 1975);
Ann. Phys. \textbf{34}, 591 (1911).
\bibitem{saito} N. Saito, J. Phys. Soc. Japan \textbf{5}, 4
(1950).
\bibitem{fixman} J.M. Peterson and M. Fixman, J. Chem. Phys. \textbf{39},
2516 (1963).
\bibitem{batchelor} G.K. Batchelor and J.T. Green, J. Fluid Mech. \textbf{56},
401 (1972).
\bibitem{batchelor2} G.K. Batchelor, J. Fluid Mech. \textbf{83},
97 (1977).
\bibitem{mazur-bedo} P. Mazur and D. Bedeaux, Physica \textbf{76}, 235
(1974).
\bibitem{bedeaux} D. Bedeaux, R. Kapral, and P. Mazur, Physica \textbf{88A}, 88
(1977).
\bibitem{saarloos} C.W.J. Beenakker, W. van Saarloos, and P. Mazur,
Physica \textbf{127A}, 451 (1984).
\bibitem{beenakker} C.W.J. Beenakker, Physica \textbf{128A}, 48
(1984).
\bibitem{russel} W.B. Russel and A.P. Gast, J. Chem. Phys. \textbf{84}, 1815 (1986).
\bibitem{brady} G. Bosis, J.F. Brady, J. Chem. Phys. \textbf{91},
1866 (1989).
\bibitem{vanderwerff1} J.C. van der Werff, C.G. de Kruiff, C. Blom, J.
Mellema, Phys. Rev. A \textbf{39}, 795 (1989).
\bibitem{vanderwerff2} J.C. van der Werff and C.G. de Kruiff, J. Rheol. \textbf{33}, 421 (1989).
\bibitem{mellema} J. Mellema, C.G. de Kruiff, C. Blom, A. Vrij,
Rheol. Acta \textbf{26}, 40 (1987).
\bibitem{verberg} R. Verberg, I.M. de Scheper, and E.G.D. Cohen, Phys. Rev. E \textbf{55}, 3143 (1997).
\bibitem{lionberger} R.A. Lionberger and W.B. Russel, Adv. Chem. Phys. \textbf{111}, 399 (2000).
\bibitem{dames} B. Dames, B.R. Morrison, N. Willembacher, Rheol. Acta \textbf{40}, 434 (2001).
\bibitem{krieger} I.M. Krieger and T.J. Dougherty, Trans. Soc. Rheol. \textbf{3}, 137 (1959).
\bibitem{quemada} D. Quemada, Rheol. Acta \textbf{16}, 82 (1977).
\bibitem{pal2} R. Pal and E. Rhodes, J. Rheol. \textbf{33}, 1021 (1989).
\bibitem{larson} R.G. Larson, \textit{The Structure and Rheology of Complex Fluids}  (Oxford, New York, 1999).
\bibitem{vandeven} T.G.M. van de Ven, \textit{Colloidal Hydrodynamics}  (Academic Press, London, 1989).
\bibitem{bruggeman} D.A.G. Bruggeman, Ann. Phys. (Leipzig) \textbf{24}, 636 (1935).
\bibitem{fuchs} K. Ghosh and R. Fuchs, Phys. Rev. B \textbf{44}, 7330 (1991); R. Fuchs and K. Ghosh, Physica A \textbf{207}, 185 (1994).
\bibitem{chang} R.L. Chang, H.P. Chiang, P.T. Leung, \emph{et al}, Solid State Communications \textbf{133}, 315 (2005).
\bibitem{mendoza} I.O. Sosa, C.I. Mendoza, and R.G. Barrera, Phys. Rev. B \textbf{63}, 144201 (2001).
\bibitem{arrhenius} Z. Arrhenius, Biochem. J. \textbf{11}, 112 (1917).
\bibitem{ball} R. Ball and P. Richmond, J. Phys. Chem. Liquids \textbf{9}, 99 (1980).
\bibitem{cichocki1} B. Cichocki, B.U. Felderhof, Phys Rev. A \textbf{43}, 5405
(1991).
\bibitem{cichocki2} B. Cichocki, B.U. Felderhof, Phys Rev. A \textbf{46}, 7723
(1992).
\bibitem{ISHPRE} I. Santamar\'{\i}a-Holek, D. Reguera and J.M. Rub\'{\i},
Phys. Rev. E \textbf{63} 051106 (2001).
\bibitem{brady04} G. Subramanian and J.F. Brady, Physica A \textbf{334}, 343 (2004).
\bibitem{rosalio} R.F. Rodr\'{\i}guez, E. Sal\'{\i}nas-Rodr\'{\i}guez, and J.W. Dufty, J. Stat. Phys. \textbf{32}, 279 (1983).
\bibitem{drossinos} Y. Drossinos, M.W. Reeks, Phys. Rev. E \textbf{71}, 031113 (2005).
\bibitem {boon} J.P. Boon, S. Yip,{\it Molecular Hydrodynamics}, (Dover, New York, 1991).
\bibitem {brady2000} J.F. Brady, J. Rheol. \textbf{44}, 629 (2000).
\bibitem{materiales} S.I. Hern\'andez, I. Santamar\'ia-Holek, C.I. Mendoza, L.F. del Castillo, Phys. Rev. E \textbf{74}, 051401 (2006).
\bibitem{SteadyState} I. Santamar\'{\i}a-Holek, J.M. Rub\'{\i} and A. P\'{e}%
rez-Madrid, New J. Phys. \textbf{7}, 35 (2005).
\bibitem{ryskin}G. Rysikin, Phys. Rev. Lett. \textbf{61}, 1442 (1988).
\bibitem{degroot} S.R. de Groot, P. Mazur, {\it Non-equilibrium
Thermodynamics} (Dover, New York, 1984).
\bibitem{landau} L.D. Landau and E.M. Lifshitz, \emph{Course of Theoretical Physics, Fluid Mechanics} (Pergammon, New York 1980), Vol. 6.
\bibitem{bhave} A.V. Bhave, R.C. Armstrong and R.A. Brown, J. Chem. Phys.
\textbf{95}, 2988 (1991).
\bibitem{de Kruif} C. G. de Kruif, E. M. F. van Iersel, A. Vrij, and W. B. Russel, J. Chem. Phys. \textbf{83}, 4717 (1986).
\bibitem{saunders} F.L. Saunders, J. Colloid. Sci. \textbf{16}, 13 (1961).
\bibitem{krieger2} I.M. Krieger, Advan. Colloid Interface Sci. \textbf{3}, 111 (1972).
\bibitem{kops} M.M. Kops-Werkhoven and H.M. Fijnaut, J. Chem. Phys. \textbf{77}, 2242 (1982).
\bibitem{bedeaux1} D. Bedeaux, J. Colloid Interface Sci. \textbf{118}, 80 (1987).
\bibitem{zhu} J.X. Zhu, D.J. Durian, J. Muller, D.A. Weitz, and D.J. Pine, Phys. Rev. Lett. \textbf{68}, 2559 (1992).
\bibitem{jones} D.A.R. Jones, B. Leary, and D.V. Boger, J. Colloid Interface Sci. \textbf{147}, 479 (1991); \textbf{150}, 84 (1992)
\bibitem{papir} Y.S. Papir and I.M. Krieger, J. Colloid Interface Sci. \textbf{34}, 126 (1970).
\end{references}
\end{document}